# A web application prototype for the multiscale modelling of seismic input

**Franco Vaccari**

Department of Mathematics and Geosciences, University of Trieste, Trieste, Italy

The Abdus Salam International Centre for Theoretical Physics, SAND Group, Trieste, Italy

E-mail: vaccari@units.it

**Abstract** A web application prototype is described, aimed at the generation of synthetic seismograms for user-defined earthquake models. The web application graphical user interface hides the complexity of the underlying computational engine, which is the outcome of the continuous evolution of sophisticated computer codes, some of which saw the light back in the middle '80s. With the web application, even the non-experts can produce ground shaking scenarios at the local or regional scale in very short times, depending on the complexity of the adopted source and medium models, without the need of a deep knowledge of the physics of the earthquake phenomenon. Actually, it may even allow neophytes to get some basic education in the field of seismology and seismic engineering, due to the simplified intuitive experimental approach to the matter. One of the most powerful features made available to the users is indeed the capability of executing quick parametric tests in near real-time, to explore the relations between each model's parameter and the resulting ground motion scenario. The synthetic seismograms generated through the web application can be used by civil engineers for the design of new seismo-resistant structures, or to analyse the performance of the existing ones under seismic load.

## Introduction

In the framework of the cooperation Project "Definition of seismic hazard scenarios and microzoning by means of Indo-European e-infrastructures", funded by Regione autonoma Friuli Venezia Giulia (Italy), a web application prototype has been developed, that enables scientists to compute a wide set of synthetic seismograms, dealing efficiently with the variety and complexity of the potential earthquake sources and of the medium travelled by the seismic waves.

The computational engine of the web application is based on the NDSHA (neo-deterministic seismic hazard assessment) methodologies (Panza et al., 2001; Panza et al., 2012) for the generation of synthetic seismograms. It allows for a rapid definition of seismic and tsunami hazard scenarios for a given event, at local or regional scales. Neo-deterministic means scenario-based methods for seismic hazard analysis, where realistic and duly validated synthetic time series, accounting for source, propagation, and site effects, are used to construct the earthquake scenarios. The user interface has been designed so to hide the intricacy of the underlying computational engine, yet it allows power users to act even on the deeper aspects of the model parameterisation.

Due to the employment of highly optimised computational codes, the web application is designed to run even locally on a properly configured laptop computer. For massive parametric tests, or single tasks targeted at modelling ground shaking scenarios at very large scale, it can be interfaced with different computational platforms, ranging from Grid computing infrastructures to HPC dedicated clusters up to Cloud computing.

The web application is addressed to engineers, urbanists, administrators, insurance companies and stake-holders who are interested in reliable territorial planning and in the design and construction of buildings and infrastructures in seismic areas. It also constitutes a powerful educational tool for seismologists and seismic engineers willing to better understand the earthquake phenomenon, since quick parametric studies can be easily performed for determining the influence of each model element on the resulting ground shaking scenarios.

## The computational engine

The basic requirement for the generation of synthetic seismograms is a proper modelling of the earthquake source properties, and of wave propagation in the medium between the source and the site of interest.

### *Source models*

The source is introduced in the medium as a discontinuity in the displacement and shear stress fields. The source function, describing the discontinuity of the displacement across the fault, is approximated with a step in time and a point in space. Imposing the continuity of the normal stress across the fault, for the representation theorem the equivalent body force in an unfaulted medium is a double-couple with null total moment (Maruyama, 1963; Burridge and Knopoff, 1964). The simplest source model adopted for the computation of the synthetic seismograms is a double-couple size scaled point source (SSPS) (Gusev, 1983). Roughly speaking, this approximation is acceptable when modelling the ground motion at large epicentral distances due to moderate magnitude events, neglecting the contribution of frequencies higher than 1 Hz.

The system is now being updated so that more realistic source models can be considered, taking into account source finiteness and the details of the rupturing process that lead to the directivity effects often observed at the sites around the fault. The most comprehensive approach being implemented is based on the extended source model (ES) described by Gusev and Pavlov (2006) and Gusev (2011), where the full characteristics of the slip distribution, and the rupturing velocity across the fault area, are taken into account. In a simplified representation, namely the size and time scaled point source model (STSPS), a combination of extended (ES) and point sources is used, with the purpose of reducing the computational times while still keeping into consideration the main characteristics of the rupturing process. In this approximation, the subsource time functions (i.e. the ones described for the ES model) are summed to obtain the equivalent single source representative of the entire space and time structure of the ES and the related Green's function. The resulting function is then convoluted with the unscaled seismogram obtained in the point source approximation.

## *Laterally homogeneous layered models*

The computational engine upon which the web application (WA hereafter) is based has its roots in the middle '80s, when the modal summation technique for the generation of P-SV synthetic seismograms (radial and vertical component of motion) was developed (Panza, 1985). Eq. 1 describes the asymptotic expression of the Fourier transform of the displacement for the radial ($u_x$) and the vertical ($u_z$) components of motion at a distance r from the source.

$$u_x^R(r,z,\omega)=\sum_{m=1}^{\infty}\frac{e^{-i3\pi/4}}{\sqrt{8\pi\omega}}\frac{e^{-ik_m^R r}}{\sqrt{r}}\frac{\left(\chi_m^R(h_S,\omega)\right)}{\sqrt{c_m^R v_m^R I_m^R}}\frac{\left(F_x^R(z,\omega)\right)}{\sqrt{v_m^R I_m^R}}$$

$$u_z^R(r,z,\omega)=\sum_{m=1}^{\infty}\frac{e^{-i\pi/4}}{\sqrt{8\pi\omega}}\frac{e^{-ik_m^R r}}{\sqrt{r}}\frac{\left(\chi_m^R(h_S,\omega)\right)}{\sqrt{c_m^R v_m^R I_m^R}}\frac{\left(F_z^R(z,\omega)\right)}{\sqrt{v_m^R I_m^R}} \tag{1}$$

The equivalent codes for the generation of the SH seismograms (transverse component of motion $u_y$, as shown in Eq. 2) were created by Florsch et al. (1991) and finally allowed for the generation of three-component seismograms.

$$u_y^L(r,z,\omega) = \sum_{m=1}^{\infty} \frac{e^{-i3\pi/4}}{\sqrt{8\pi\omega}} \frac{e^{-ik_m^L r}}{\sqrt{r}} \frac{\left(\chi_m^L(h_s,\omega)\right)}{\sqrt{c_m^L v_m^L I_m^L}} \frac{\left(F_y^L(z,\omega)\right)}{\sqrt{v_m^L I_m^L}} \tag{2}$$

At that time, given the limited (when compared to nowadays...) computational facilities available to the researchers, algorithm efficiency and proper computer code implementation was of uttermost importance. The beauty of the modal summation technique computer codes for layered, inelastic structural models, based on the original paper by Panza, lies in the separation of the calculation of the spectral quantities related with the medium properties (phase velocity $c$, group velocity $v$, energy integral $I$ and complex (due to anelasticity) wavenumber $k$ appearing in Eqs. 1 and 2), from those that describe the source and its position with respect to the sites of interest (radiation pattern $\chi$, epicentral distance $r$, rupturing model etc). In such a way, once the medium is described by defining each layer's thickness, density, *Vp* and *Vs* wave velocities and the *Qp* and *Qs* factors that define attenuation, the "lengthy" computation of the spectral quantities that do not depend on the source position can be performed just once, and the obtained quantities can simply be stored and reused for as many seismograms have to be computed.

The "lengthy" adjective is really rooted in the '80s. At that time, as reported by Panza (1985), the computation of the spectral quantities associated with a layered model required about one hour of CPU time on an IBM 370/168 computer. In the author's memories, it was even longer than that. Due to job queueing and a less powerful mainframe computer, jobs had to be submitted in the evening and results were obtained, with some luck, the day after. Now, the same results can be obtained on a modern laptop in a couple of seconds. Most obviously because of the improved performance of modern computers, and partly due to code optimisation applied since then, and I/O rearrangements allowed by the increased memory now available to programs.

A further reduction in the computational times is achieved when a sequence of seismograms is generated keeping constant the hypocentral depth. Thus, the terms describing the eigenfunctions calculated at the depth of the source (terms $F_x$ and $F_z$ in Eq. 1 and $F_y$ in Eq. 2) can be computed for the first seismogram only, and reused for the following ones. Therefore, even in the '80s, after the spectral quantities were obtained, the modal summation approach allowed for a "quick" generation of the time series: 300 s for the first seismogram, and 30 s for the others, when keeping constant the hypocentral depth. Come back onboard of a modern laptop, and a thousand seismograms can be generated in a couple of seconds.

While the layered model approximation cannot be considered satisfactory for every scenario in the world, nevertheless the modal summation technique is an unbelievably powerful tool to explore in near real-time the influence of single model (source and layer) parameters on the generated ground shaking scenarios. It can help to quickly constrain part of the model parameterisation for the optimised planning of more complicated and computationally intensive tasks. It is also at the base of the seismic hazard computations at regional scale, as discussed in detail by Panza et al (2001) and Panza et al (2012).

## *Laterally heterogeneous structural models*

More sophisticated computational approaches can be used when the geological setting of the studied area can hardly be approximated by a flat layered model (Panza et al, 2001). For the WA, the hybrid technique developed originally by Fäh and Panza (1994) is the engine adopted for the computation of synthetic seismograms along 2D heterogeneous profiles. A laterally homogeneous inelastic layered model is defined to represent the average lithospheric properties along the path from the source to the vicinity of the local, heterogeneous structure of interest. In this part of the model wave propagation is modelled by the modal summation technique, according to what described in the previous Section. So there is no time penalty in this part of the model, associated with the length of the path. The generated wavefield is then introduced in the mesh that defines the local heterogeneous area characterising the site of interest, where it is propagated according to the finite-difference scheme. With this hybrid approach, source, path, and site effects are all taken into account, and detailed ground shaking scenarios can be efficiently evaluated along the 2D profile even at large distances from the epicentre.

The computational time spent for the modelling depends primarily on the size of the finite difference mesh, and on the duration requested for the seismograms. Again, with respect to the the middle ’90s, when the theory and codes were developed, execution time has greatly improved. Results that required days of mainframe computer CPU time, can now be obtained in a matter of tens of minutes to hours, depending on the model characteristics.

The hybrid method has demonstrated its validity as a tool to perform the seismic microzonation of urban areas (e.g. Panza et al, 2002; Alvarez et al, 2005; Harbi et al, 2007; Zuccolo et al, 2008; Amponsah et al, 2009; Mohanty et al, 2013). Given its widespread application, quite some resources have been dedicated to the development of helper applications that permit an easy and nearly error-proof construction of the finite difference mesh. Still, we are quite far from reaching the level of speedy interactivity allowed by the modal summation tech-nique for laterally homogeneous models. Furthermore, some training is required for the users to self-certify the quality of the results obtained. It may happen yet, although not very often, that the model parameterisation has to be finely tuned by the user, when the default input configuration fails to produce acceptable results.

Other computational techniques based on the modal summation theory may simplify and speed up the generation of ground shaking scenarios in laterally heterogeneous media. Their field of applicability is mostly driven by the characteristics of the medium, above all the geometric properties of the heterogeneities, and the acoustic impedance contrast across the interfaces. For instance, when the lateral heterogeneity can be reasonably approximated by a sharp vertical boundary separating two layered quarter-spaces, the modal coupling technique (Vaccari et al, 1989; Romanelli et al, 1997) can be efficiently used. In this approach, the energy carried by the incoming modes, excited by the source, is transmitted across the boundary and redistributed in the modes that characterise the second structure. This technique remains valid also in the presence of a sequence of vertical boundaries, noting that its effectiveness is inversely proportional to the number of considered interfaces: when their number increases, the ramification of the cross-coupling paths between modes of different order may grow dramatically, as well as the need to consider multiple reflections between the boundaries.

When considering smoothly varying media, a different approach can be taken for the computation of the synthetic seismograms. A good choice is the so called WKBJ approximation (acronym of the names Wentzel, Kramers, Brillouin and Jeffreys). Here, the heterogeneity can be seen as a perturbation of an initial lateral homogeneous model and, if such a perturbation is small within a wavelength, a procedure based on the ray method can be used to construct an approximate solution corresponding to the wave field (see e.g. Woodhouse, 1974; Yanovskaya, 1989).

The computational package based on the programs by Panza (1985) and Florsch et al (1991) has been already expanded and integrated with new codes to permit the computation of synthetic seismograms in 2D and 3D heterogeneous media (La Mura, 2009; La Mura et al, 2011; Panza et al, 2012). The interfacing with the WA is in progress, as it requires the development of some additional routines not yet available, mostly aimed at the simplification of the model preparation; the same is true for the mode coupling approach.

## *Future developments*

In general, the WA is so structured, that it poses no big problems in implementing new computational techniques, should they became available, capable of pro-ducing ground shaking scenarios better and faster for specific classes of structural and source models. The computational engine is well separated from the user interface (written in the Xojo cross-platform, object-oriented language), so in principle any software developed in any programming language could be installed on the dedicated number-crunching hardware. When adding a new modelling tool, the extra work implied is limited to the modification of an existing panel or the addition of a new one to the WA user interface, properly designed to allow an easy definition of the input model, and the visualisation of the computational results. This generally requires the development of some middleware that improves and simplify the process of input preparation, and validates the data provided by the user.

## The web application user interface

### *Login page*

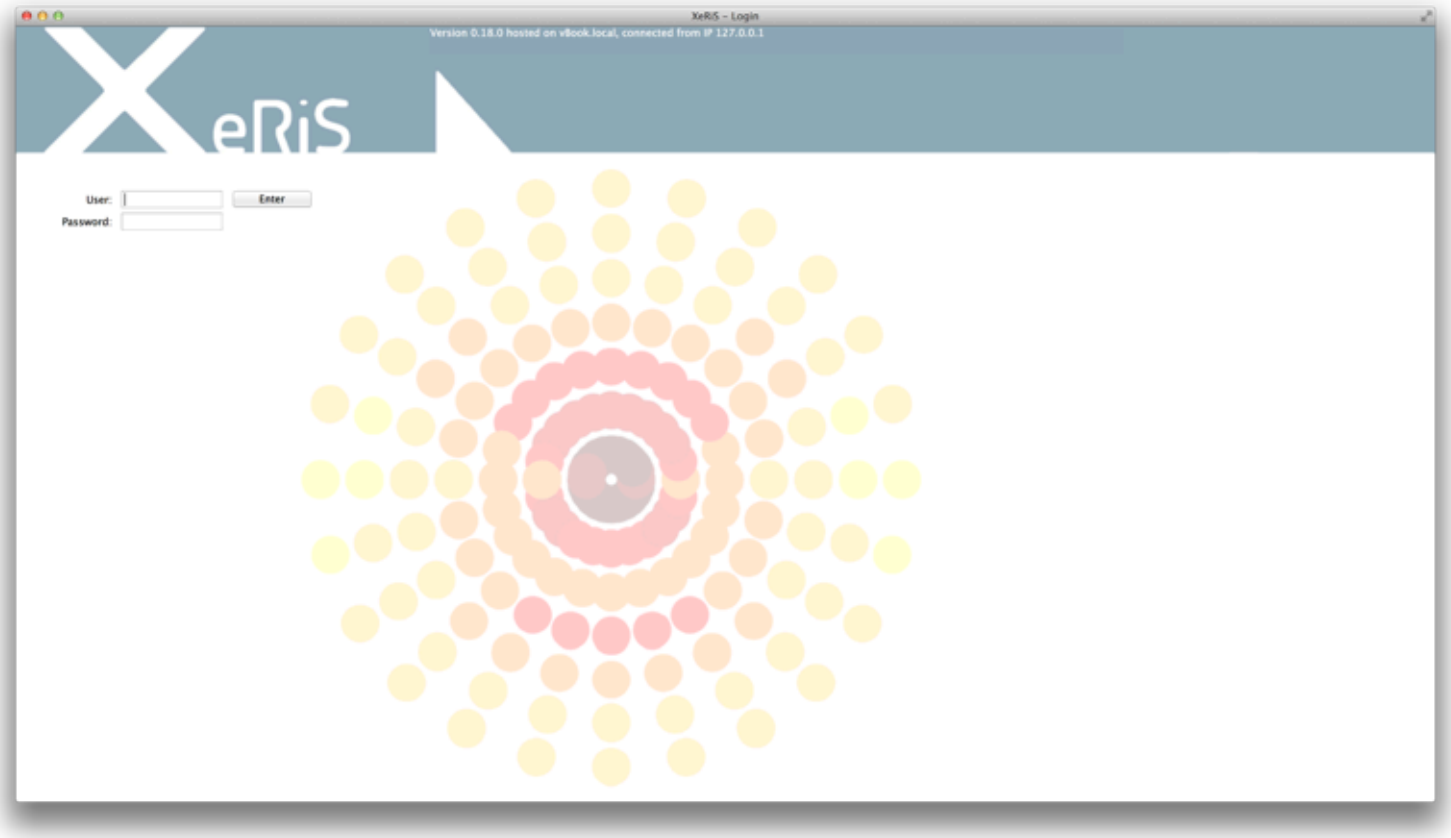

**Fig. 1** Login page of the web application.

The WA requires the user to pass through a login page, shown in Fig. 1. Each of the user's input data and computational results are stored into well separated areas of the file system. In no way a logged-in user can access other users' data navigating through the WA interface panels. The administrator of the WA can create new accounts, lock or delete users and clean their stored computations when needed, authorise access to a subset of functionalities on a per-user basis, and check the system status.

After entering the system, a tabbed interface keeps the different computational tasks well separated. In the following, each tab functionality is described, with some screenshots shown to better highlight the way the user interacts with the system. Each panel has a help page for the basic explanation of the available functionalities, with a list of bibliographic references useful for those interested in a deeper understanding of the physical phenomenon and of the computational tools used for its modelling.

## *Laterally homogeneous structural models*

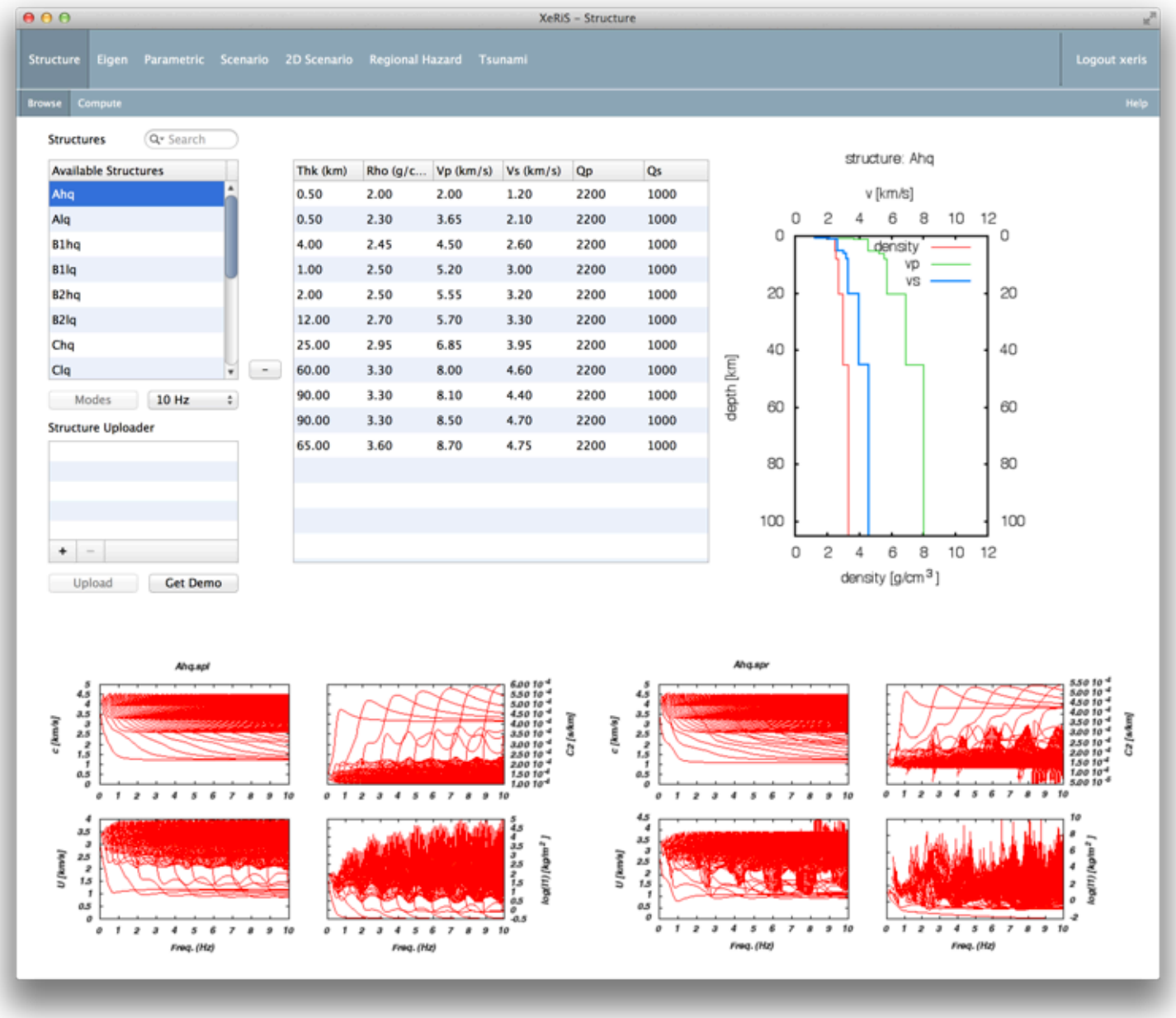

**Fig. 2** Panel dedicated to the definition of the laterally homogeneous layered models, and the related frequency domain quantities required to compute the synthetic seismograms by the modal summation technique. Top, from left to right: the list of structural models available to the user, with the uploader that allows the addition of more, user-defined structures; the properties associated with the layers of the selected structure; the plot of the model with density, *Vp* and *Vs* depicted. Bottom: the spectral quantities calculated for Love and Rayleigh modes.

The first panel available to the user after logging into the system is the one dedicated to the computation of the frequency quantities associated with the layered structural models (Fig. 2).

By default, some standard models are made available to the user upon the first login, so that one can immediately start experimenting with the WA. The models correspond to classes of ground types mentioned in seismic design rules like Eurocode 8 (EN 1998-1, 2004), ranging from hard rocks of type "A" to soft soils of type "E". For each class of ground, two variations are prepared with different quality factors $Qs = 1000$ for slowly attenuating models, and $Qs = 100$ for structures characterised by higher attenuation. $Qp$ is taken equal to $2.2 \cdot Qs$.

The user must prepare a simple text file with the description of the layers as shown in Fig. 3, and then upload the file through the dedicated control. After that, the model appears in the list of available structures and can be selected to generate its Rayleigh and Love modes with the choice of the cutoff frequency for the computations: 1 Hz for modelling the ground motion very far from the source, or 10 Hz for obtaining ground shaking scenarios at shorter distances.

| thk | rho | Vp | Vs | Qp | Qs |
|---|---|---|---|---|---|
| 2.000 | 2.20 | 4.67 | 2.70 | 220 | 100 |
| 3.000 | 2.40 | 5.54 | 3.20 | 220 | 100 |
| 3.000 | 2.50 | 5.89 | 3.40 | 220 | 100 |
| 7.000 | 2.70 | 5.97 | 3.45 | 330 | 150 |
| 4.000 | 2.75 | 6.23 | 3.60 | 330 | 150 |
| 9.000 | 2.80 | 6.40 | 3.70 | 330 | 150 |
| 12.000 | 2.90 | 6.92 | 4.00 | 330 | 150 |
| 5.000 | 3.10 | 7.09 | 4.10 | 330 | 150 |
| 11.000 | 3.24 | 7.27 | 4.20 | 330 | 150 |
| 9.000 | 3.25 | 7.44 | 4.30 | 330 | 150 |
| 9.000 | 3.26 | 7.61 | 4.40 | 330 | 150 |
| 18.000 | 3.29 | 7.79 | 4.50 | 330 | 150 |

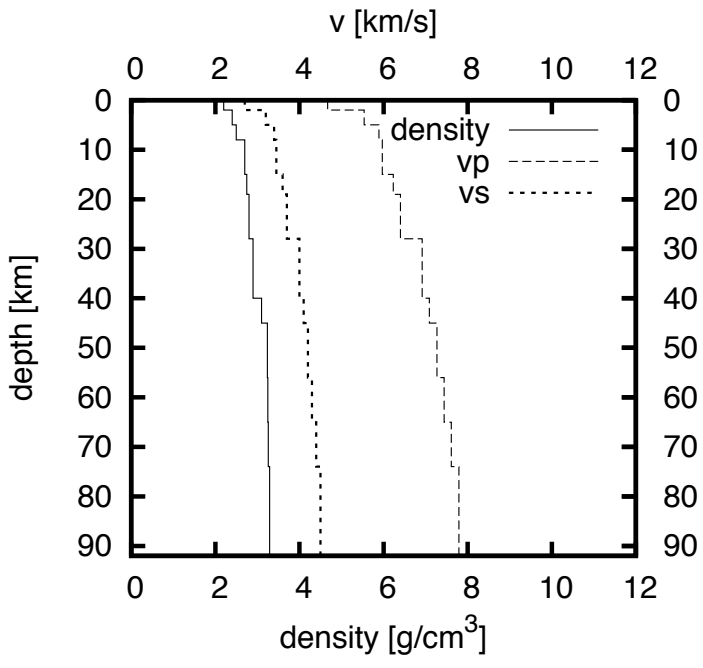

**Fig. 3** Example of the content of a text file (left) the user should prepare for uploading a layered structural model (right) to the web application. Thickness (thk) is given in km, density (rho) in g/cm$^3$, layer velocities (*Vp* and *Vs*) in km/s while attenuations for *P* and *S* velocity (*Qp* and *Qs* respectively) are adimensional quantities. In this examples, the *Vp* value has been obtained from the *Vs* in the hypothesis of a Poissonian medium, but there is actually no constrain on the *Vp*/*Vs* ratio that can be adopted for each layer.

The underlying computational engine takes care of splitting the physical layers defined by the user into thinner layers that satisfy the requirements of the algorithms at the base of the modal summation technique, to avoid potential overflow conditions during the evaluation of the matrix elements for any given layer (Panza, 1985). The inclusion of low-velocity layers is also permitted by the methodology.

While not too long, still the user has to wait some time before the computation get finalised, and the modes are shown in the web application panel of Fig. 2. The amount of this delay mostly depends on the system load at the time of the job submission. In general, it should not exceed a couple of minutes. Once the computational request has been submitted to the system, the user interface signals the job status through the appearance of an icon in the toolbar bar. The colour of the icon is associated with the job status: grey if the job is queued for execution; yellow after the execution has started, green when the modes are finally available and can be used for the generation of the synthetic seismograms. While the computations are taking place, there is no lock in the user interface. Other panels can still be accessed to play with the separate functionalities of the WA.

It may sometimes happen that for whatever reason, usually associated with typing errors in the file that defines the layer properties, or with very peculiar characteristics of the structural model, the mode generation fails. This condition is signalled by a red colour in the toolbar icon. In such a case, the user should first check if some errors have been made in the preparation of the structural model, or modify slightly the layer properties to try to overcome the problem. The management of error conditions has not yet been thoroughly considered at this prototypal stage, but will definitely get the attention it deserves as the WA approaches a more mature state.

## *Eigenfunction visualisation*

This panel has to be taken mostly as an educational tool, rather than something designed or required for the definition of a ground shaking scenario. It allows the user to explore the distribution with depth of the eigenfunctions (displacements and stresses) associated with a given layered model. The user can choose the structural model for which the eigenfunctions should be generated, and visualise their depth distribution for selected Love and Rayleigh modes, at a given frequency. In such a way, one can understand which modes and which frequencies will be better excited by a source placed at a given depth, therefore contributing to the construction of the synthetic seismogram. For instance, the example of Fig. 4 shows that in order to excite mode n. 4 at the frequency of 2 Hz a source should be placed no deeper than 6 km.

The evaluation of the eigenfunctions is truly instantaneous, so the user can compute and visualise them in a really interactive way, quickly getting familiar with their behaviour as a function of the structure characteristics, the mode index or the frequency considered. Contrary to all other panels, due to the speedy computation, the job is not queued upon submittal but is executed immediately at the push of the dedicated button. Therefore no toolbar icon is required in this panel to signal the job status. Only the structures for which the modes have been already generated appear in the list of the selectable models.

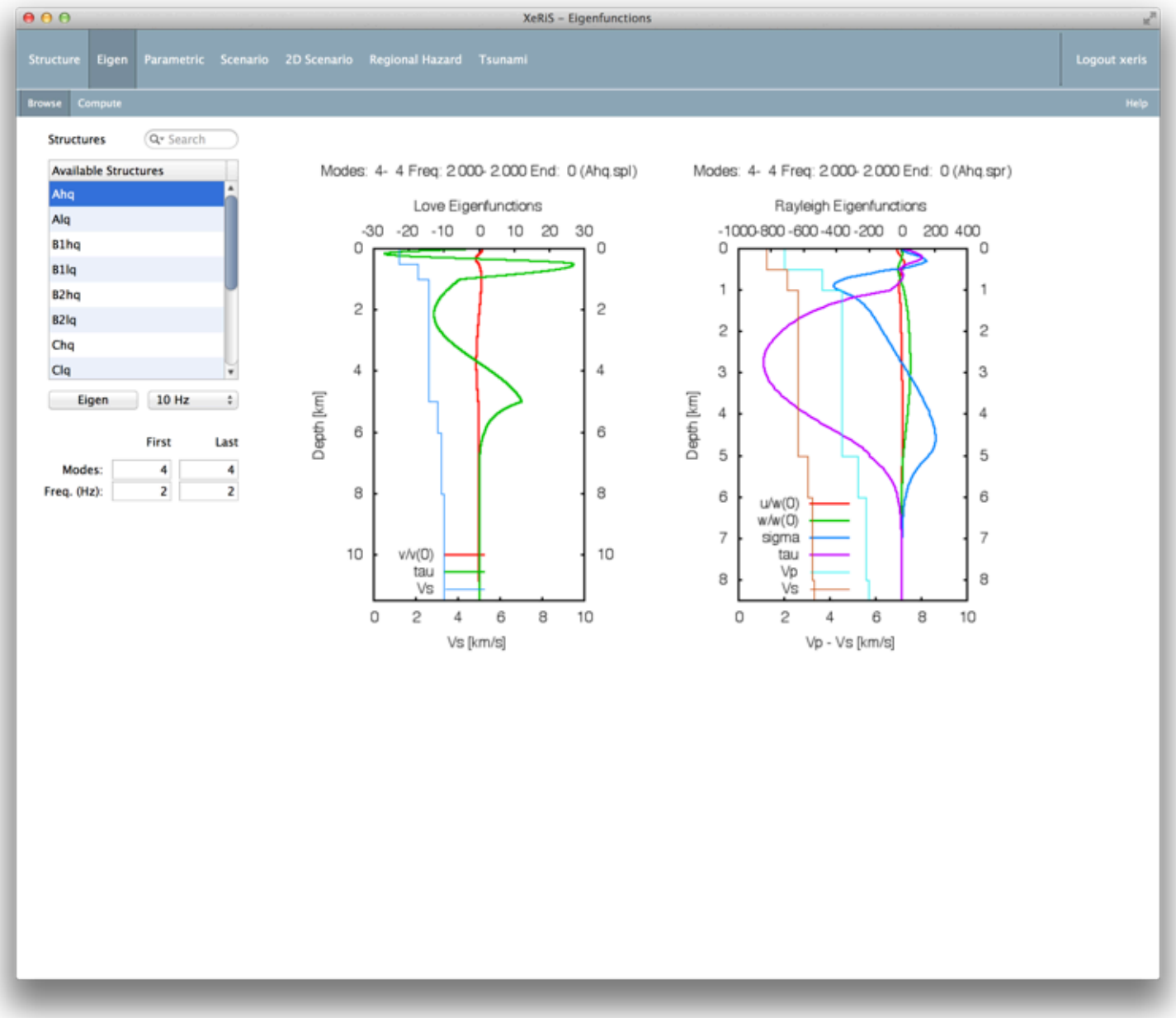

**Fig. 4** Panel dedicated to the computation of the eigenfunctions (displacement and stresses) associated with a structural model for a given mode and frequency.

## *Parametric tests*

This panel is dedicated to the quick execution of parametric tests, where a set of synthetic seismograms is generated based on the variation of a single source parameter. The user can choose to variate in a loop one of the following quantities: event magnitude, epicentral distance, hypocentral depth, fault dip and rake, and the azimuthal position of the observation point with respect to the fault strike. The time series (radial, transverse and vertical component of motion) are generated in a matter of seconds, with the choice of visualising ground displacement, velocity or acceleration. A summarising graph showing the peak values taken from the seismograms as a function of the varied parameter is also given, to intuitively demonstrate the dependency of the ground motion amplitude on the single source characteristic being analysed. The example of epicentral distance variation is shown in Fig. 5.

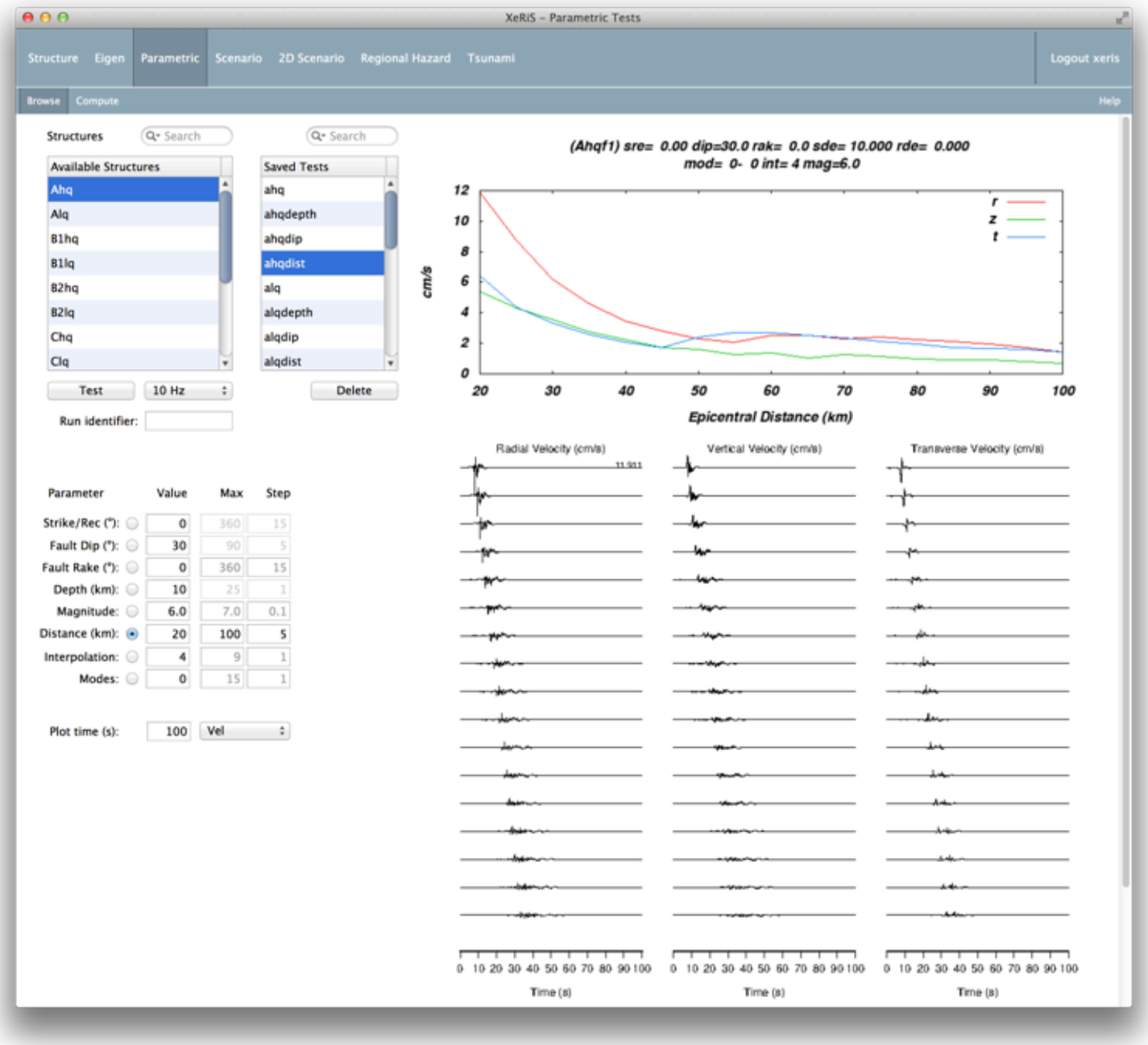

**Fig. 5** Panel dedicated to the computation of a set of three-component synthetic seismograms obtained varying a single model parameter, to grasp its influence on the obtained ground shaking. Top, from left to right: the list of available structural models; the list of stored parametric tests; the graph summarising the dependency of the peak ground motion on the selected parameter (epicentral distance in the shown example). Bottom: the definition of the source model, with the selected parameter and its range of variation; the plot of the three component seismograms.

A list of the structures for which the modes have been already computed is presented in the panel. This allows to easily explore the dependency of the ground shaking also on the layer properties. Each experiment can be quickly repeated selecting the desired structural model from the list. The results obtained with each simulation are properly labelled and stored for later retrieval and comparison with other experiments. Stored results can of course be deleted by the user when they are no longer needed.

For a given source parameterisation, and mostly for educational purposes, it is also possible to visualise a set of time series, each generated for a single mode of the considered structure. For instance, after configuring a loop over modes running from index 1 to 10, the contributions coming from the fundamental mode (index=1) up to the ninth higher mode (index=10) will be separately shown in the seismogram plot. Furthermore, if the starting index of the mode loop is set to 0 instead of 1, the first seismogram shown will be the “complete” one, that is the one obtained summing up all the modes associated with the considered structure.

## *Earthquake scenarios*

The next functionality of the WA is targeted at the generation of full earthquake scenarios, by calculating synthetic seismograms all around the epicentre in a user-defined range of epicentral distances and with properly discretised azimuthal and distance steps. In this modelling, the computational time depends on the number of seismograms that must be generated. But unless the user requires the investigation of a very broad area at very small discretisation steps, it is still a matter of seconds to obtain the requested scenario (Fig. 6).

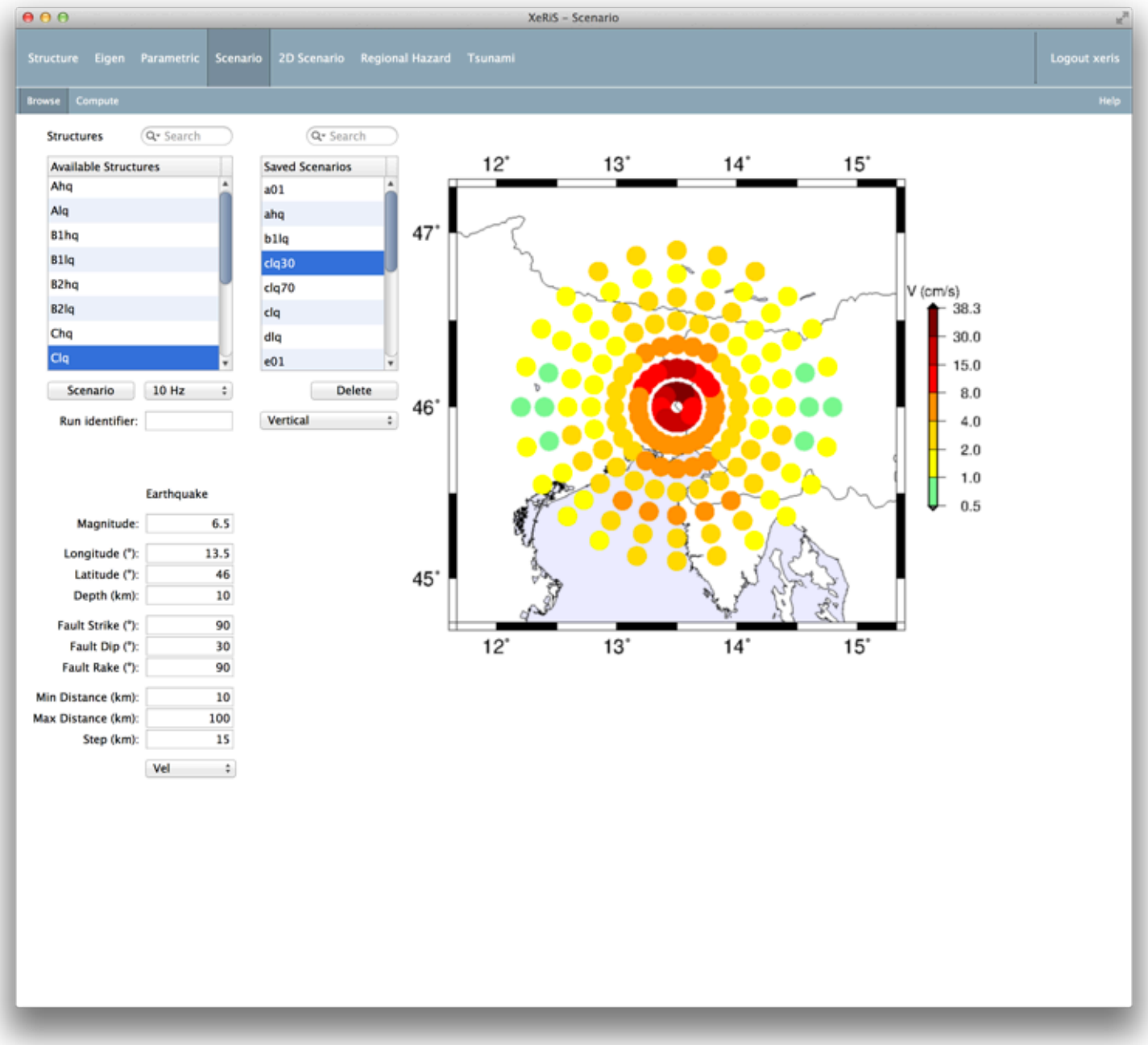

**Fig. 6** Panel dedicated to the computation of ground shaking scenarios for a single earthquake. Top, from left to right: the list of available structural models for the selected cutoff frequency; the list of stored scenarios with the popup menu for the selection of the component to be visualised; the map of ground motion shaking (vertical PGV in the shown example). Bottom left: the parameters adopted to define the scenario.

As in the case of the computation of modes, already described, also for the generation of earthquake scenarios a toolbar icon appears after a computational job has been submitted, with the icon's colour signalling the job status.

The computed scenarios remain available to the user, and can be retrieved at any moment by selecting its associated label in the dedicated listbox.

## *Earthquake scenarios along heterogeneous profiles*

The hybrid technique used for the computation of ground shaking scenarios along laterally heterogeneous profiles requires a much larger amount of CPU time with respect to the modelling procedures described up to now. It is definitely not the kind of process that can be done interactively. To simplify the operations, and to keep the user interface clean, the dedicated panel has been further organised in sub-panels.

The first one is dedicated to the definition of the model, and to the submission of new jobs (Fig. 7). The user can choose the layered structural model, among those for which the modes have been already computed, that represents the average lithospheric properties along the path from the source to the vicinity of the sites of interest. After that, the characteristics of the heterogeneous profile that includes those sites must be selected.

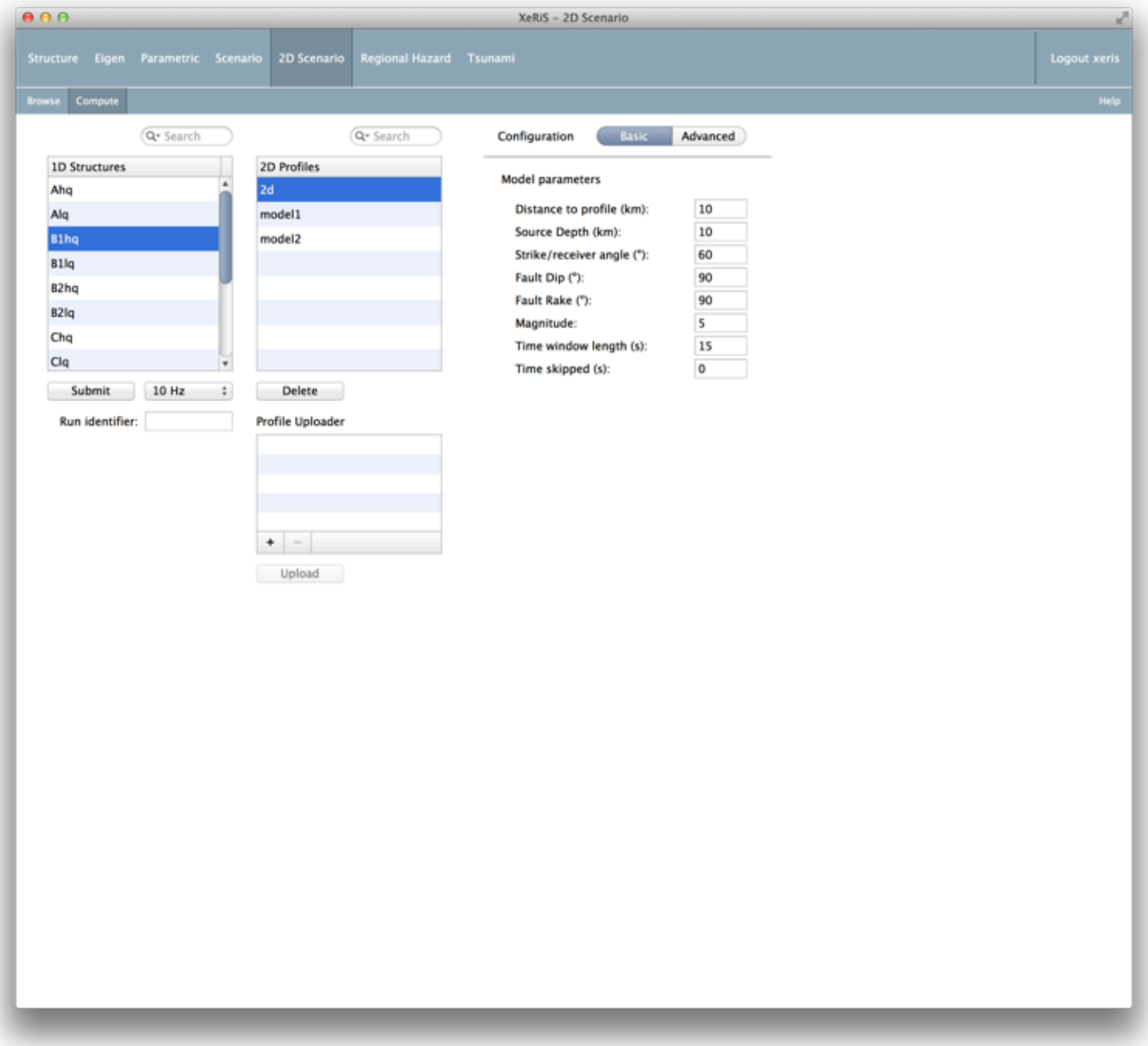

**Fig. 7** Panel dedicated to the computation of ground shaking scenarios along a laterally heterogeneous profile. On the left is the list of available structural models used to define the average lithospheric properties along the path from the source to the site of interest, for which the modes have been computed already. In the middle is the list of available heterogeneous profiles already loaded into the WA. On the right there are further parameters, mostly source-related, that the user must adjust, in “Basic” or “Advanced” mode, to properly configure the experiment.

Finally, in the “Basic” configuration mode, the source parameterisation must be done. An “Advanced” mode is eventually made available to the power users, that allows for a fine tuning of the finite difference model. Tweaking to the mesh discretisation, and modifications to other rather obscure technical parameters, usually kept out of reach to the unexperienced users, can be applied here.

After all the above has been properly defined, either in “Basic” or “Advanced” mode, the modelling can be started at the push of a button.

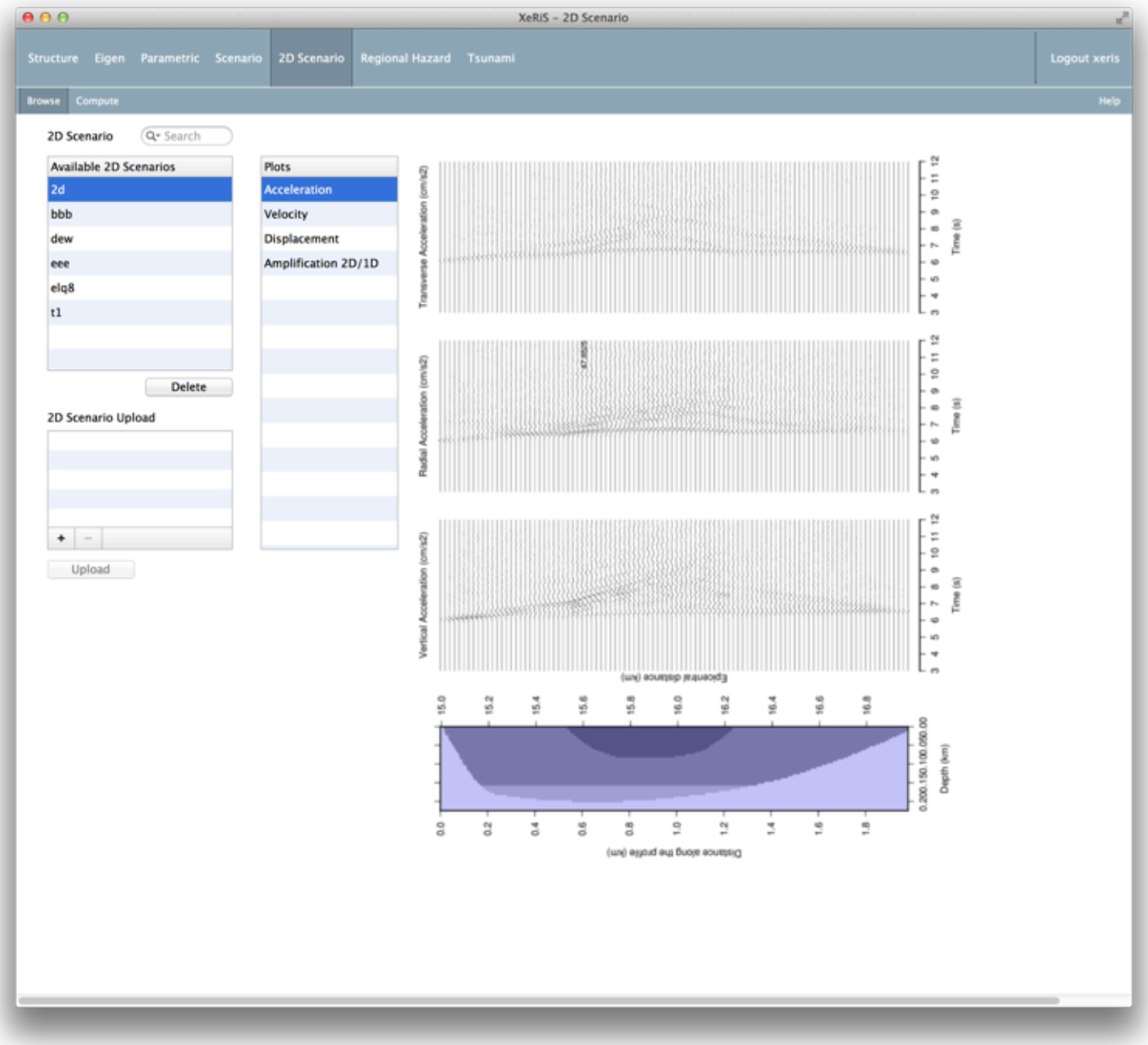

**Fig. 8** Three component synthetic seismograms computed along the heterogeneous profile.

The second panel (Fig. 8 and Fig. 9) is organised for the retrieval of the scenarios that have been already computed. No need for the user to remain logged in while the computational task proceeds. When a job finishes, its associated label is automatically added to the dedicated listbox. At the next login, if the job has completed, the user will be able to select the related entry in the list of available scenarios, and visualise the results produced. The choices are between the three components synthetic seismograms (in acceleration, velocity or displacement), shown in Fig. 8, and the spectral amplifications computed with respect to the average lithospheric model, taken as reference, as shown in Fig. 9.

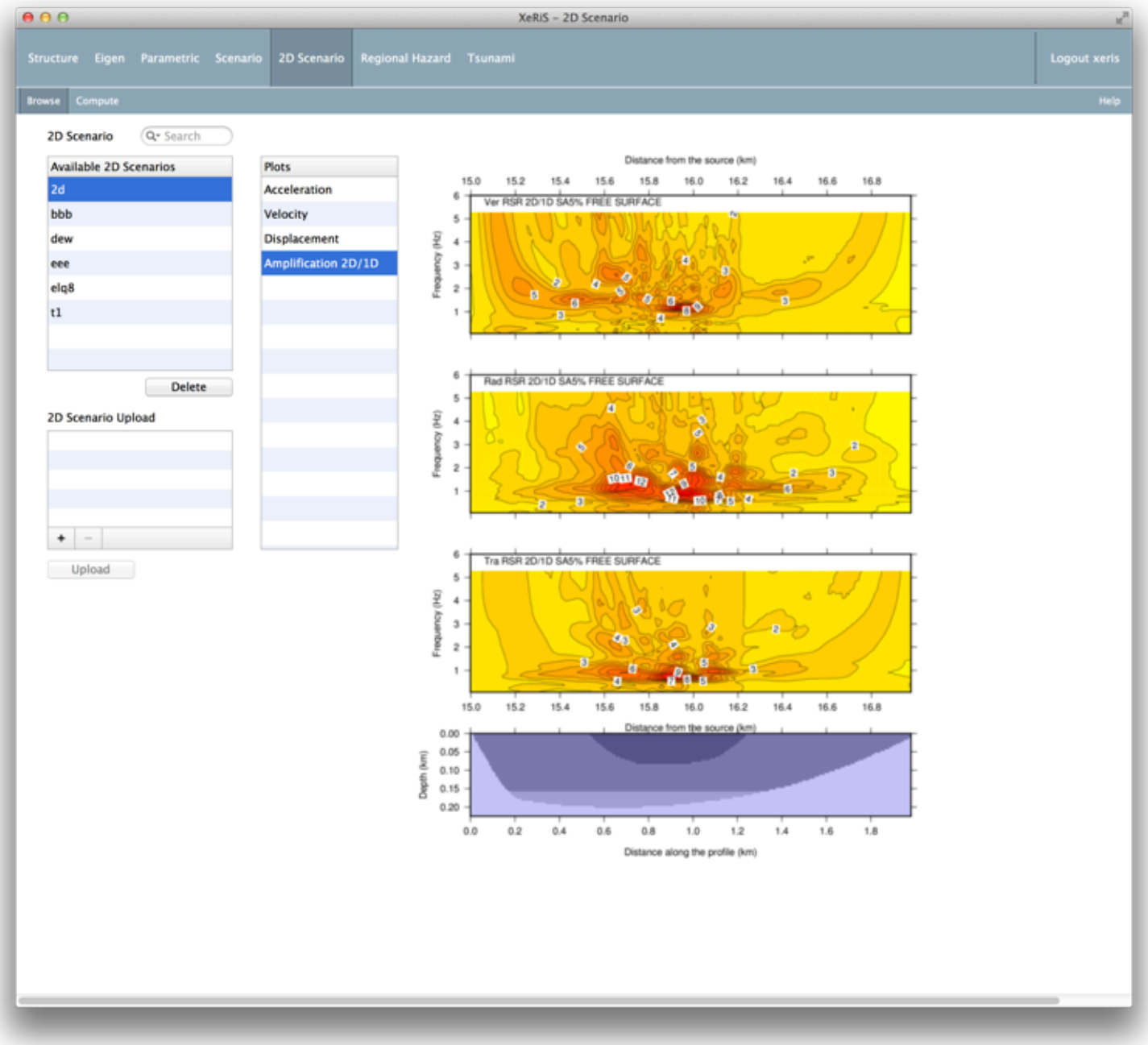

**Fig. 9** Spectral amplifications computed along the heterogeneous profile.

A little more must be told about the preparation of the laterally heterogeneous part of the model. This step once required the manual editing of a quite complicated file, in which the lateral heterogeneities were defined through the superposition of rectangular “patches”, each covering a part of the layered model representing the average lithospheric structure. Every patch is characterised by its own set of density, *P* and *S* waves velocities and attenuation values. For complicated geometries the editing process was lengthy, convoluted and extremely error-prone. Most often, a stupid mistake in the preparation of this file resulted in wrong scenarios produced after days of wasted CPU time. Such a computational nightmare is now relegated into the past. A dedicated desktop application has been developed (Fig. 10), that allows the user to draw the model using the standard tools available in painting programs (straight and freehand pencils, colour picker, filler, magnifier lens etc), plus some dedicated tools for the calibration of the model size, the digitisation of points and polygons coordinates, and for the chromatic correction of the model if an image acquired with a scanner is used as a reference when designing the profile. Each colour used in drawing the model is associated with a unique set of layer properties, that can be stored in a library for later reuse. After the drawing is complete, a rasterisation algorithm produces the rectangular patches required by the computational engine, and exports them in a properly formatted file, as needed by the finite difference program.

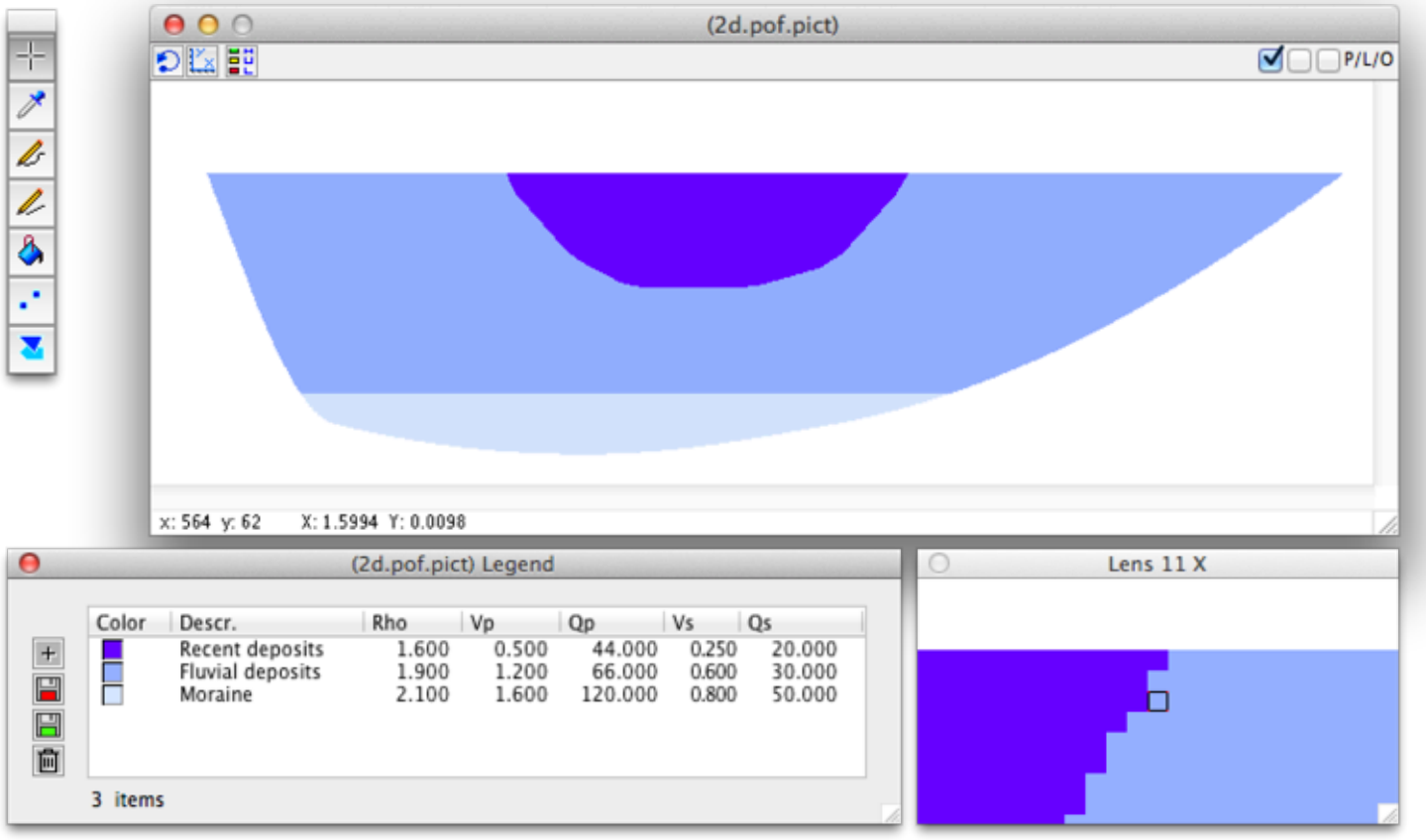

**Fig. 10** The main elements of the desktop application developed for the preparation of the laterally heterogeneous profiles. From the top-left corner, clockwise: the tools palette, the main window, the magnified view around the cursor position and the legend with the layer properties.

## *Seismic hazard scenarios at regional scale*

The definition of seismic hazard with the neo-deterministic approach, as fully described in Panza et al (2001) and Panza et al (2012), requires the preparation of several input datasets. They are needed for the definition of the characteristics (magnitude and focal mechanism) of the potential earthquake sources distributed within the active seismogenic zones, and for the description of the average lithospheric properties of the Earth in the considered region.

Once the input files are prepared, the computer package performs the following tasks, with no further user intervention:

- discretisation into cells of size 0.2° x 0.2° of the seismicity data taken from the historical earthquake catalogue, assigning to each cell the maximum magnitude obtained into each cell after a smoothing window has been applied to the original data;
- definition of a representative focal mechanism to be associated to all the sources belonging to each seismogenic zone;
- definition of a grid of sites evenly distributed in the considered region;

- generation of the paths between sources and sites, properly sorted by structural model and then by source depth, in view of an optimised computation of the synthetic seismograms;
- generation of the three-components synthetic seismograms (displacements, velocities, accelerations) for all the paths previously identified, with the possibility to adopt the SSPS or STSPS models;
- extraction and mapping of the peak values on the grid of sites.

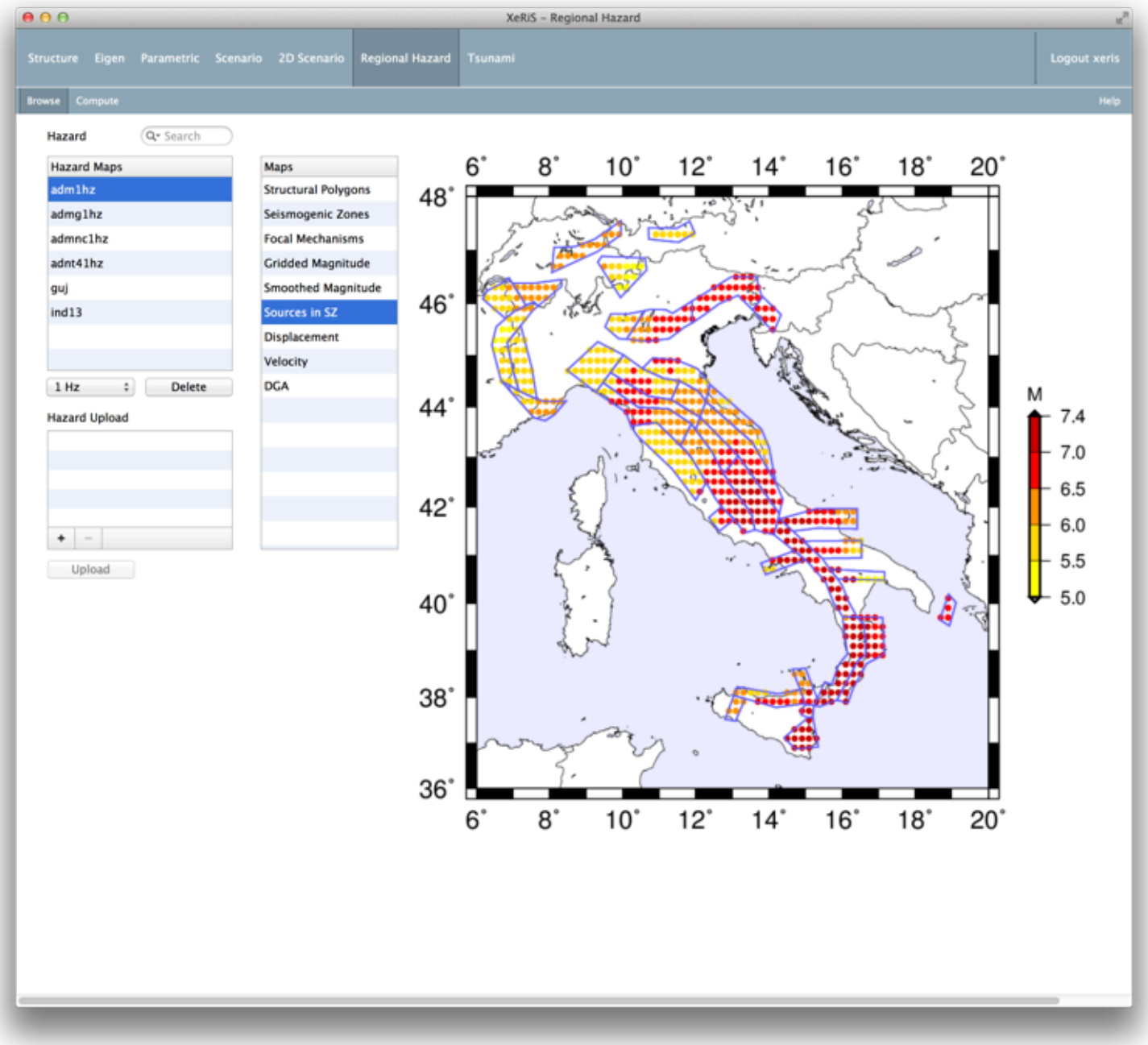

**Fig. 11** Panel of the WA dedicated to the browsing of the input data and of the corresponding seismic hazard scenarios. From left to right: the list of available computational experiments, with the control that permits to upload more items; the list of available maps for the selected experiment; the map. Here the distribution of sources inside the seismogenic zones, and their magnitude, is shown.

The computational engine is based on the modal summation technique. Depending on the extension of the area being studied, and on the range of epicentral distances, requested by the user or automatically chosen by the system according to the magnitude of the considered sources, it might be necessary to compute a huge amount of seismograms. Typical numbers may span from a few hundred (for small areas defined ad-hoc for quick preliminary parametric tests), to millions of seismograms for large sub-continental areas. For the cited extremes, the computational time required varies roughly between few seconds and about a day.

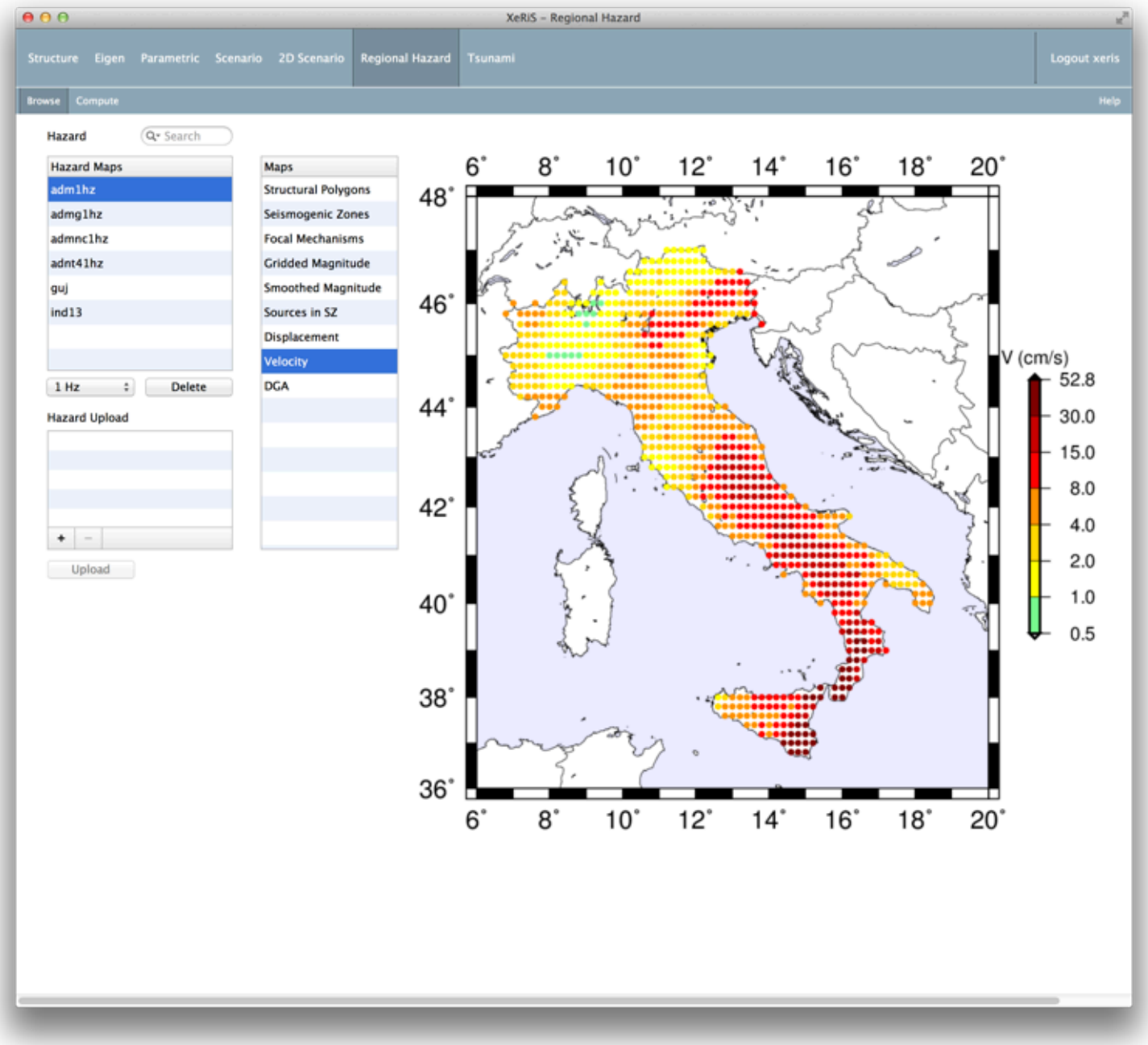

**Fig. 12** In the same panel shown in Fig. 11, the obtained PGV distribution is visualised here.

At the current stage of development, the "Regional Hazard" panel is fully functional in its browsing capabilities, as shown in Fig. 11. and Fig. 12. In that sub-panel, the user can upload the compressed archives generated automatically by the package at the end of the execution. Each archive contains both the input data and the output maps representing the hazard scenario.

For the generation of new scenarios from within the web application, the dedicated sub-panel layout is almost ready, but not all the planned functionalities are currently implemented. At present, it permits the upload of the required input datasets, including the optional ones that may be considered by the package and not yet mentioned here: the seismogenic nodes obtained through the morphostructural analysis as defined by Gorshkov et al (2002; 2004; 2009), and the alerted areas declared by the CN or M8 algorithms (Peresan et al, 2005). What is currently being developed, and must be considered a truly essential improvement, is a set of helper applications aimed at guaranteeing that the uploaded input files are, at the very least, properly formatted according to the guidelines of the package. Better yet, a further feature is already planned, that will allow a quality check of the data, targeted at the identification of unreasonable values mistakenly introduced in the datasets.

## *Tsunami hazard scenarios*

This panel is the latest addition to the WA. Similarly to the seismic hazard panel, it currently allows the browsing of the uploaded results of tsunami modelling (Fig. 13). The computational engine upon which the modelling is based is the one described in Panza et al (2000). The capability of generating new scenarios through a dedicated and user friendly web interface is currently being implemented, and will be shortly made available to the beta testers of the WA.

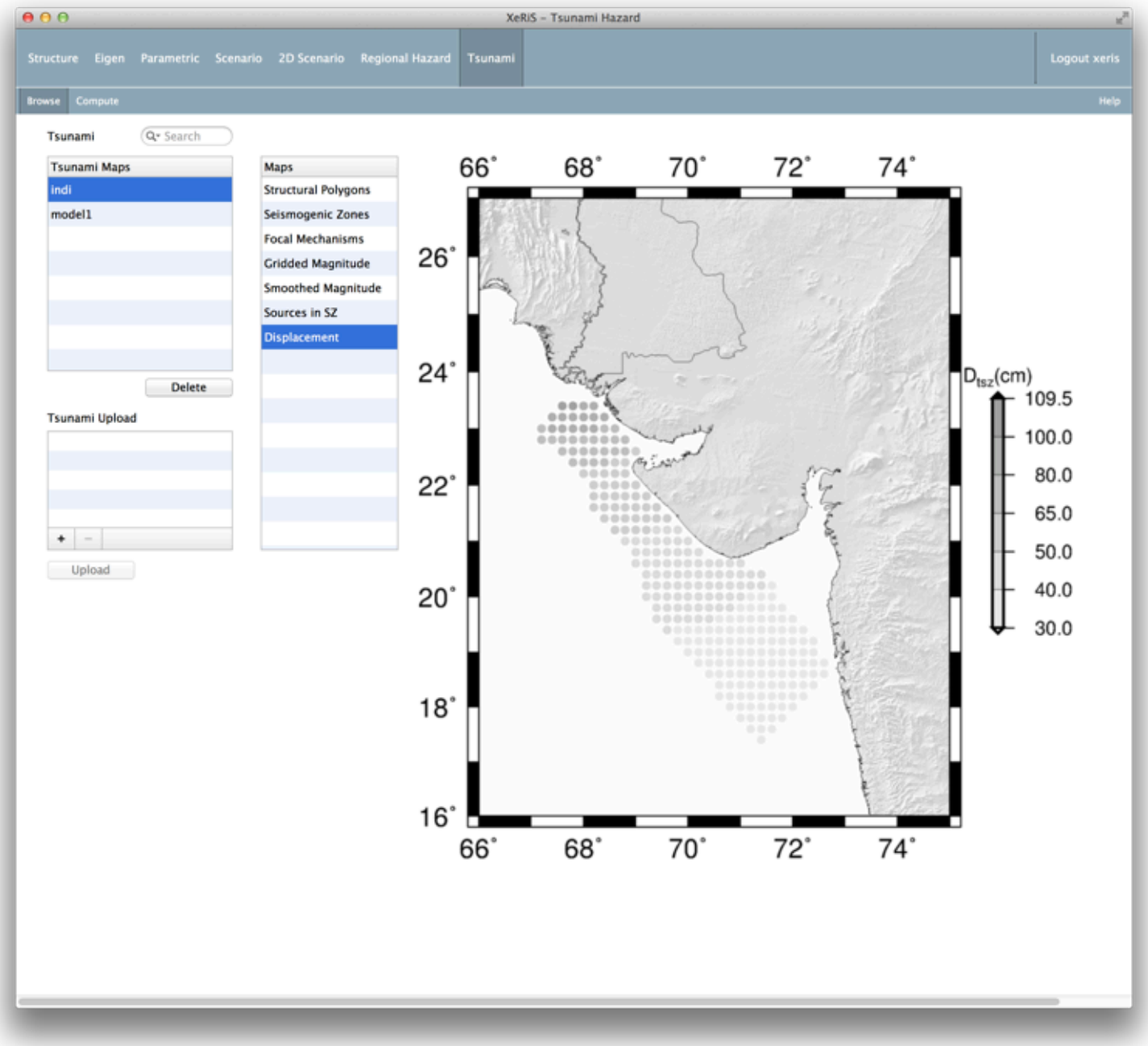

**Fig. 13** Panel dedicated to the management of tsunami hazard calculations. Here the height of the modelled tsunami wave is shown.

## Final remarks

It is a matter of fact that the seismic regulations currently adopted worldwide are mostly based on probabilistic estimates of the seismic hazard. This is very likely due to the lack of adequate tools that might have provided viable alternatives at the time the regulations were formulated.

The web application prototype here described aims at demonstrating that something new can finally be done to improve the preparedness against future events. With the underlying NDSHA computational engine, the end-user can be kept well isolated from the complexity of the modelling tools, and there is no steep learning curve for him to fight against, before he can generate his first ground shaking scenario. This is particularly true for the design and execution of quick parametric tests in laterally homogeneous layered models, where the source or layer properties are varied, and the effects (or lack thereof) of their variation on the ground shaking can be immediately verified in the obtained scenarios. Under these conditions, it is really a matter of minutes for civil engineers or city planners to get a first-order estimate of the seismic input expected for the next earthquake.

Conversely, the execution of massive experiments configured to explore the whole model parameter space, within the limits of the current knowledge, or the adoption of more realistic heterogeneous models, can be programmed taking advantage of modern computer architectures, based on the notions of Grid and Cloud computing.

The future development of this web application prototype is planned in such a way that the choice of the computational hardware on which the modelling will be executed should be almost transparent to the user. It has to become as simple as checking a radio button on the graphical user interface.

The benefits of the methodology upon which the web application is based have been recently proved in the city of Trieste (Italy). Commissioned by the Provincia di Trieste Authorities, ground shaking scenarios have been generated in the city along several 2D profiles. The amplifications obtained due to the combined interaction of source, path and site effects have lead to spectral accelerations higher than those predicted by the official hazard maps, based on probabilistic studies and currently adopted by the law. The seismic input specifically computed at selected sites of interest for the Provincia di Trieste Authorities has been used to verify the behaviour of some relevant buildings under the seismic load, and to plan the retrofitting that will hopefully avoid heavy damaging, or even the collapse of the structures when the earthquake will hit.

**Acknowledgments** First and foremost, this web application couldn't have seen the light without the underlying computational engine, based on the pioneering work by Prof. Panza. Giuliano's energy, and his continuos push and encouragement have played a decisive role in shaping up and transforming the original codes into the friendly tools they now are.

The web application development saw the light within the project "Definition of seismic hazard scenarios and microzoning by means of Indo-European e-infrastructures" funded by Regione autonoma Friuli Venezia Giulia in the framework of the interventions aimed at promoting, at regional and local level, the cooperation activities for development and international partnership - "Progetti Quadro ai sensi della Legge regionale n. 19 del 30 ottobre 2000". I'm truly grateful to Antonella for her impeccable project management.

The recent, strong evolution of the underlying computational engine would have never been possible without the dedicated work by the friends who made my life better here at the University: Fabio, Elisa, Andrea, Cristina and Davide, to name just those more deeply involved in the coding. And of course Enrico, for he knows what...

Stefano and Francesco have introduced me into the new universe of grid and cloud computing. I really enjoyed it, and I do hope our collaboration will continue and expand in the future.

Finally, a deep gratitude goes to my beloved family. I wish all the computational time spared through algorithm and code optimisation might be magically converted into time spent with them. Apparently I'm not too good in doing so, but don't loose your faith, I'll keep trying... :-)